\documentstyle[amssymb,preprint,prb,aps]{revtex}

\begin{document}

\begin{titlepage}

\title
{Quantum nucleation in ferromagnets with tetragonal and hexagonal symmetries}

\author{Yi Zhou, Rong L\"{u},
Jia-Lin Zhu, and Lee Chang} 
\address{Center for Advanced Study, Tsinghua University,
Beijing 100084, P. R. China}
\date{\today}
\maketitle
\begin{abstract}
The phenomenon of quantum nucleation is studied 
in a ferromagnet in the presence of a
magnetic field at an arbitrary angle.
We consider the magnetocrystalline anisotropy with tetragonal
symmetry and that with hexagonal symmetry, respectively.
By applying the instanton method in the
spin-coherent-state path-integral representation, we
calculate the dependence of the rate of quantum nucleation
and the crossover temperature on the orientation and strength of the
field for a thin film and for a bulk solid.
Our results show that the rate of quantum nucleation and the crossover
temperature depend on the 
orientation of the external magnetic
field distinctly, which provides a possible experimental
test for quantum nucleation in
nanometer-scale  ferromagnets.

\noindent
{\bf PACS number(s)}:  75.45.+j, 73.40.Gk, 75.30.Gw, 75.50.Gg 
\end{abstract}

\end{titlepage}

\section*{I. Introduction}

The tunneling of macroscopic object, known as Macroscopic Quantum Tunneling
(MQT), is one of the most fascinating phenomena in condensed matter physics.
In the last decade, the problem of quantum tunneling of magnetization in
nanometer-scale magnets has attracted a great deal of theoretical and
experimental interest.\cite{1} MQT in magnetic systems are interesting from
a fundamental point of view as they can extend our understanding of the
transition from quantum to classical behavior. On the other hand, these
phenomena are important to the reliability of small magnetic units in memory
devices and the designing of quantum computers in the future. And the
measurement of magnetic MQT quantities such as the tunneling rates could
provide independent information about microscopic parameters such as the
magnetocrystalline anisotropies and the exchange constants. All this makes
magnetic quantum tunneling an exciting area for theoretical research and a
challenging experimental problem.

The problem of quantum nucleation of a stable phase from a metastable one in
ferromagnetic films is an interesting fundamental problem which allows
direct comparison between theory and experiment.\cite{2} Consider a
ferromagnetic film with its plane perpendicular to the easy axis determined
by the magnetocrystalline anisotropy energy depending on the crystal
symmetry. A magnetic field {\bf H} is applied in a direction between
perpendicular and opposite to the initial easy direction of the
magnetization {\bf M}, which favors the reversal of the magnetization. The
reversal occurs via the nucleation of a critical bubble, which then the
nucleus does not collapse, but grows unrestrictedly in volume. If the
temperature is sufficiently high, the nucleation of a bubble is a thermal
overbarrier process, and the rate of thermal nucleation follows the
Arrhenius law $\Gamma _T\varpropto \exp \left( -U/k_BT\right) $, with $k_B$
being the Boltzmann constant and $U$ being the height of energy barrier. In
the limit of $T\rightarrow 0$, the nucleation is purely quantum-mechanical
and the rate goes as $\Gamma _Q\varpropto \exp \left( -{\cal S}_{cl}/\hbar
\right) $, with ${\cal S}_{cl}$ being the classical action or the WKB\
exponent which is independent of temperature. Because of the exponential
dependence of the thermal rate on $T$, the temperature $T_c$ characterizing
the crossover from quantum to thermal regime can be estimated as $%
k_BT_c=\hbar U/{\cal S}_{cl}$.

The problem of quantum nucleation was studied by Privorotskii\cite{3} who
estimated the exponent in the rate of quantum nucleation based on the
dimensional analysis. Chudnovsky and Gunther\cite{4} studied the quantum
nucleation of a thin ferromagnetic film in the presence of an external
magnetic field along the opposite direction to the easy axis at zero
temperature by applying the instanton method in the spin-coherent-state
path-integral representation. Ferrera and Chudnovsky extended the quantum
nucleation to a finite temperature.\cite{5} Kim studied the quantum
nucleation in a thin ferromagnetic film placed in a magnetic field at an
arbitrary angle.\cite{6}

It is noted that the previous results\cite{4,5,6} of quantum nucleation were
obtained for ferromagnetic sample with the simplest form of the
magnetocrystalline anisotropy energy such as the biaxial symmetry, and the
model considered in Refs. 4 and 5 was confined to the condition that the
magnetic field be applied along the opposite direction to the easy axis. The
purpose of this paper is to extend the previous results of quantum
nucleation in ferromagnetic system with simple biaxial symmetry to that of
system with a more general symmetry, such as tetragonal and hexagonal
symmetry. The generic quantum nucleation problem, however, and the easiest
to implement in practice, is that of ferromagnets with a general structure
of magnetocrystalline anisotropy in a magnetic field applied at a some angle 
$\theta _H$ to the anisotropy axis. This problem does not possess any
symmetry and for that reason is more difficult mathematically. It is worth
pursuing, however, because of its significance for experiments.\cite{1} In
this paper the magnetic field is applied in an arbitrary direction between
perpendicular and opposite to the initial easy axis ($\widehat{z}$ axis).
Our interest in studying quantum nucleation of magnetic bubbles with a more
general structure of magnetocrystalline anisotropy in an arbitrarily
directed magnetic field is stimulated by the fact that the corresponding
experiment would be most easy to perform and to interpret. Within the
instanton approach, we present the numerical results for the WKB exponent in
quantum nucleation of a thin ferromagnetic film with the magnetic field
applied in a range of angles $\pi /2<\theta _H<\pi $, where $\theta _H$ is
the angle between the initial easy axis ($\widehat{z}$ axis) and the field.
We also discuss the $\theta _H$ dependence of the crossover temperature $T_c$
from purely quantum nucleation to thermally assisted processes. Our results
show that the distinct angular dependence, together with the dependence of
the WKB\ exponent on the strength of the external magnetic field, may
provide an independent experimental test for quantum nucleation in a
ferromagnetic film. Quantum nucleation (the description involves space-time
instantons), being a field theory problem, is more difficult than tunneling
of magnetization in single-domain particles, both at the conceptual and at
the technical level. Therefore, this paper provides {\it a nontrivial
generalization of uniform rotation of magnetization vector (homogeneous spin
tunneling) in single-domain magnets\cite{13,15} to a nonuniform rotation of
magnetization in bulk magnets with a more general structure of
magnetocrystalline anisotropy in the presence of a magnetic field at an
arbitrary angle}. Compared with the tunneling in single-domain particles, a
local tunneling event in a bulk magnet can trigger instability on a much
greater scale, which leads to really macroscopic consequences.\cite{1} In
experiments, it may be easier to monitor single nucleation events in a thin
film than to detect the magnetization reversal in a nanometer-scale
particle. Therefore, our theoretical results for a general structure of
magnetocrystalline anisotropy in an arbitrarily directed field will be more
applicable for experimental tests of quantum nucleation. Besides the
importance from the fundamental point of view, processes of quantum
nucleation and collapse of magnetic bubbles are potentially important for
quantum limitations on the density and long-term reliability of the data
storage in magnetic memory devices and designing of quantum computer.

This paper is structured in the following way. In Sec. II, we review briefly
some basic ideas of quantum nucleation of magnetization in ferromagnets. In
Secs. III and IV, we study quantum nucleation of magnetization in
ferromagnets with tetragonal and hexagonal symmetry in an external magnetic
field applied in the $ZX$ plane with a range of angles $\pi /2\leq \theta
_H<\pi $. The conclusions and discussions are presented in Sec. V.

\section*{II. The physical model}

For a spin tunneling problem, the rate of magnetization reversal by quantum
tunneling is determined by the imaginary-time transition amplitude from an
initial state $\left| i\right\rangle $ to a final state $\left|
f\right\rangle $ as 
\begin{equation}
U_{fi}=\left\langle f\right| e^{-HT}\left| i\right\rangle =\int {\cal D}%
\left\{ {\bf M}\left( {\bf r},\tau \right) \right\} \exp \left( -{\cal S}%
_E/\hbar \right) ,  \eqnum{1}
\end{equation}
where ${\cal S}_E$ is the Euclidean action which includes the Euclidean
Lagrangian density ${\cal L}_E$ as 
\begin{equation}
{\cal S}_E=\int d\tau d^3{\bf r}{\cal L}_E.  \eqnum{2}
\end{equation}
For ferromagnets at sufficiently low temperature, all the spins are locked
together by the strong exchange interaction, and therefore only the
orientation of magnetization ${\bf M}\left( {\bf r},\tau \right) $ can
change but not its absolute value. For that reason the field ${\bf M}\left( 
{\bf r},\tau \right) $ is equivalent to the fields $\theta \left( {\bf r}%
,\tau \right) $ and $\phi \left( {\bf r},\tau \right) $, which are spherical
coordinates of ${\bf M}$. In this case the measure of the path integral $%
{\cal D}\left\{ {\bf M}\left( {\bf r},\tau \right) \right\} $ in Eq. (1) is
equivalent to 
\begin{equation}
\int {\cal D}\left\{ \theta \left( {\bf r},\tau \right) \right\} {\cal D}%
\left\{ \phi \left( {\bf r},\tau \right) \right\} =\lim_{\varepsilon
\rightarrow 0}\prod_{k=1}^N\left( \frac{2S+1}{4\pi }\right) \sin \theta
_kd\theta _kd\phi _k,  \eqnum{3}
\end{equation}
where $\varepsilon =\max \left( \tau _{k+1}-\tau _k\right) $ and $%
S=M_0/\hbar \gamma $ is the total spin of ferromagnet. Here $\gamma $ is the
gyromagnetic ratio and $M_0$ is the magnitude of magnetization.

In the spin-coherent-state representation, the magnetic Lagrangian is given
by 
\begin{equation}
{\cal L}_E=i\frac{M_0}\gamma \left( \frac{d\phi \left( {\bf r},\tau \right) 
}{d\tau }\right) \left[ 1-\cos \theta \left( {\bf r},\tau \right) \right]
+E\left( \theta ,\phi \right) .  \eqnum{4}
\end{equation}
The first term in Eq. (4) is a total imaginary-time derivative, which has no
effect on the classical equations of motion, but it is crucial for the
spin-parity effects.\cite{1,7,8,9,10,11} However, for the closed instanton
trajectory described in this paper (as shown in the following), this time
derivative gives a zero contribution to the path integral, and therefore can
be omitted.

The energy density in Eq. (4) is 
\begin{equation}
E\left( \theta ,\phi \right) =E_a\left( \theta ,\phi \right) +E_{ex}\left(
\theta ,\phi \right) ,  \eqnum{5}
\end{equation}
where $E_a$ includes the magnetocrystalline anisotropy energy and the energy
due to the external magnetic field, and $E_{ex}$ is the exchange energy 
\begin{equation}
E_{ex}=\frac \alpha 2\left( \partial _iM_j\right) ^2=\frac \alpha 2%
M_0^2\left[ \left( \nabla \theta \right) ^2+\sin ^2\theta \left( \nabla \phi
\right) ^2\right] ,  \eqnum{6}
\end{equation}
where $\alpha $ is the exchange stiffness..\cite{12} The magnetocrystalline
anisotropy energy for tetragonal and hexagonal symmetry is shown in Sec. III
and IV, respectively. In the semiclassical limit, the rate of quantum
nucleation, with an exponential accuracy, is given by 
\begin{equation}
\Gamma _Q\varpropto \exp \left[ -{\cal S}_E^{\min }/\hbar \right] , 
\eqnum{7}
\end{equation}
where ${\cal S}_E^{\min }$ is obtained along the trajectory that minimizes
the Euclidean action ${\cal S}_E$.

\section*{III. Ferromagnets with tetragonal symmetry}

In this section, we study the quantum nucleation of magnetization in
ferromagnets with tetragonal symmetry in the presence of a magnetic field at
arbitrary angles in the $ZX$ plane, which has the following
magnetocrystalline anisotropy energy 
\begin{equation}
E_a\left( \theta ,\phi \right) =K_1\sin ^2\theta +K_2\sin ^4\theta
-K_2^{\prime }\sin ^4\theta \cos \left( 4\phi \right) -M_0H_x\sin \theta
\cos \phi -M_0H_z\cos \theta ,  \eqnum{8}
\end{equation}
where $K_1$, $K_2$ and $K_2^{\prime }$ are the magnetic anisotropy
coefficients, and $K_1>0$. In the absence of the magnetic field, the easy
axes of this system are $\pm \widehat{z}$ for $K_1>0$. And the field is
applied in the $ZX$ plane at $\pi /2<\theta _H<\pi $. Then the total energy
is given by 
\begin{eqnarray}
E\left[ \theta \left( {\bf r},\tau \right) ,\phi \left( {\bf r},\tau \right)
\right] &=&K_1\sin ^2\theta +K_2\sin ^4\theta -K_2^{\prime }\sin ^4\theta
\cos \left( 4\phi \right) +\frac \alpha 2M_0^2\left[ \left( \nabla \theta
\right) ^2+\sin ^2\theta \left( \nabla \phi \right) ^2\right]  \nonumber \\
&&-M_0H_x\sin \theta \cos \phi -M_0H_z\cos \theta +E_0,  \eqnum{9}
\end{eqnarray}
where $E_0$ is a constant which makes $E\left( \theta ,\phi \right) $ zero
at the initial state. By applying the similar method in Ref. 15, we can
perform a Gaussian integration over the variable $\phi $ in the path
integral and reduce the system to that with only one variable $\delta $ (as
shown in the following). Then it is possible to perform the rest of the
calculation by using the instanton method. This method simplifies the
problem tremendously, compared to the problem where the action depended on $%
\theta \left( \tau \right) $ and $\phi \left( \tau \right) $, though a
complete mathematical equivalence to the initial problem is preserved.

By introducing the dimensionless parameters as 
\begin{equation}
\overline{K}_2=K_2/2K_1,\overline{K}_2^{\prime }=K_2^{\prime }/2K_1,%
\overline{H}_x=H_x/H_0,\overline{H}_z=H_z/H_0,  \eqnum{10}
\end{equation}
Eq. (9) can be rewritten as 
\begin{eqnarray}
\overline{E}\left( \theta ,\phi \right) &=&\frac 12\sin ^2\theta +\overline{K%
}_2\sin ^4\theta -\overline{K}_2^{\prime }\sin ^4\theta \cos \left( 4\phi
\right) +\frac{\alpha M_0^2}{4K_1}\left[ \left( \nabla \theta \right)
^2+\sin ^2\theta \left( \nabla \phi \right) ^2\right]  \nonumber \\
&&-\overline{H}_x\sin \theta \cos \phi -\overline{H}_z\cos \theta +\overline{%
E}_0,  \eqnum{11}
\end{eqnarray}
where $E\left( \theta ,\phi \right) =2K_1\overline{E}\left( \theta ,\phi
\right) $, and $H_0=2K_1/M_0$. At finite magnetic field, the plane given by $%
\phi =0$ is the easy plane, on which $\overline{E}_a\left( \theta ,\phi
\right) $ reduces to 
\begin{equation}
\overline{E}_a\left( \theta ,\phi =0\right) =\frac 12\sin ^2\theta +\left( 
\overline{K}_2-\overline{K}_2^{\prime }\right) \sin ^4\theta -\overline{H}%
\cos \left( \theta -\theta _H\right) +\overline{E}_0.  \eqnum{12}
\end{equation}
The initial angle $\theta _0$ is determined by $\left[ d\overline{E}_a\left(
\theta ,0\right) /d\theta \right] _{\theta =\theta _0}=0$, and the critical
angle $\theta _c$ and the dimensionless critical field $\overline{H}_c$ are
determined by both $\left[ d\overline{E}_a\left( \theta ,0\right) /d\theta
\right] _{\theta =\theta _c,\overline{H}=\overline{H}_c}=0$ and $\left[ d^2%
\overline{E}_a\left( \theta ,0\right) /d\theta ^2\right] _{\theta =\theta _c,%
\overline{H}=\overline{H}_c}=0$, which leads to 
\begin{eqnarray}
\frac 12\sin \left( 2\theta _0\right) +\overline{H}\sin \left( \theta
_0-\theta _H\right) +4\left( \overline{K}_2-\overline{K}_2^{\prime }\right)
\sin ^3\theta _0\cos \theta _0 &=&0,  \eqnum{13a} \\
\frac 12\sin \left( 2\theta _c\right) +\overline{H}_c\sin \left( \theta
_c-\theta _H\right) +4\left( \overline{K}_2-\overline{K}_2^{\prime }\right)
\sin ^3\theta _c\cos \theta _c &=&0,  \eqnum{13b} \\
\cos \left( 2\theta _c\right) +\overline{H}_c\cos \left( \theta _c-\theta
_H\right) +4\left( \overline{K}_2-\overline{K}_2^{\prime }\right) \left(
3\sin ^2\theta _c\cos ^2\theta _c-\sin ^4\theta _c\right) &=&0.  \eqnum{13c}
\end{eqnarray}
Assuming that $\left| \overline{K}_2-\overline{K}_2^{\prime }\right| \ll 1$,
we obtain the critical magnetic field and the critical angle as 
\begin{eqnarray}
\overline{H}_c &=&\frac 1{\left[ \left( \sin \theta _H\right) ^{2/3}+\left|
\cos \theta _H\right| ^{2/3}\right] ^{3/2}}\left[ 1+\frac{4\left( \overline{K%
}_2-\overline{K}_2^{\prime }\right) }{1+\left| \cot \theta _H\right| ^{2/3}}%
\right] ,  \eqnum{14a} \\
\sin \theta _c &=&\frac 1{\left( 1+\left| \cot \theta _H\right|
^{2/3}\right) ^{1/2}}\left[ 1+\frac 83\left( \overline{K}_2-\overline{K}%
_2^{\prime }\right) \frac{\left| \cot \theta _H\right| ^{2/3}}{1+\left| \cot
\theta _H\right| ^{2/3}}\right] .  \eqnum{14b}
\end{eqnarray}
In the low barrier limit, i.e., $\epsilon =1-\overline{H}/\overline{H}%
_c\rightarrow 0$, by using Eqs. (13b) and (13c) we obtain the approximate
equation for $\eta \left( \equiv \theta _c-\theta _0\right) $ in the order
of $\epsilon ^{3/2}$, 
\begin{eqnarray}
&&-\epsilon \overline{H}_c\sin \left( \theta _c-\theta _H\right) +\eta
^2\left[ \frac 32\overline{H}_c\sin \left( \theta _c-\theta _H\right)
+3\left( \overline{K}_2-\overline{K}_2^{\prime }\right) \sin \left( 4\theta
_c\right) \right]  \nonumber \\
&&+\eta \left\{ \epsilon \overline{H}_c\cos \left( \theta _c-\theta
_H\right) -\eta ^2\left[ \frac 12\overline{H}_c\cos \left( \theta _c-\theta
_H\right) +4\left( \overline{K}_2-\overline{K}_2^{\prime }\right) \cos
\left( 4\theta _c\right) \right] \right\} =0.  \eqnum{15}
\end{eqnarray}
Introducing $\delta \equiv \theta -\theta _0$ ($\left| \delta \right| \ll 1$
in the small $\epsilon $ limit), we derive the energy $\overline{E}\left(
\theta ,\phi \right) $ as 
\begin{eqnarray}
\overline{E}\left( \delta ,\phi \right) &=&\overline{K}_2^{\prime }\left[
1-\cos \left( 4\phi \right) \right] \sin ^4\left( \theta _0+\delta \right) +%
\overline{H}_x\left( 1-\cos \phi \right) \sin \left( \theta _0+\delta \right)
\nonumber \\
&&+\frac{\alpha M_0^2}{4K_1}\left[ \left( \nabla \theta \right) ^2+\sin
^2\theta \left( \nabla \phi \right) ^2\right] +\overline{E}_1\left( \delta
\right) ,  \eqnum{16}
\end{eqnarray}
where $\overline{E}_1\left( \delta \right) $ is a function of only $\delta $
given by 
\begin{eqnarray}
\overline{E}_1\left( \delta \right) &=&\left[ \frac 12\overline{H}_c\sin
\left( \theta _c-\theta _H\right) +\left( \overline{K}_2-\overline{K}%
_2^{\prime }\right) \sin \left( 4\theta _c\right) \right] \left( \delta
^3-3\delta ^2\eta \right)  \nonumber \\
&&+\left[ \frac 18\overline{H}_c\cos \left( \theta _c-\theta _H\right)
+\left( \overline{K}_2-\overline{K}_2^{\prime }\right) \cos \left( 4\theta
_c\right) \right] \left( \delta ^4-4\delta ^3\eta +6\delta ^2\eta ^2-4\delta
^2\epsilon \right)  \nonumber \\
&&+4\left( \overline{K}_2-\overline{K}_2^{\prime }\right) \epsilon \delta
^2\cos \left( 4\theta _c\right) .  \eqnum{17}
\end{eqnarray}

It can be shown that in the region of $\pi /2<\theta _H<\pi $, $0<\theta
_c<\pi /2$, $\eta $ and $\delta $ are of the order of $\sqrt{\epsilon }$,
the second or third term in Eq. (17) is smaller than the first term in the
small $\epsilon $ limit. It is convenient to use dimensionless variables 
\begin{equation}
{\bf r}^{\prime }=\epsilon ^{1/4}{\bf r}/r_0,\tau ^{\prime }=\epsilon
^{1/4}\omega _0\tau ,\overline{\delta }=\delta /\sqrt{\epsilon },  \eqnum{18}
\end{equation}
where $r_0=\sqrt{\frac{\alpha M_0^2}{2K_1}}$, and $\omega _0=2\gamma K_1/M_0$%
. Then the Euclidean action Eq. (2) for $\pi /2<\theta _H<\pi $ becomes 
\begin{eqnarray}
{\cal S}_E\left[ \overline{\delta }\left( {\bf r}^{\prime },\tau ^{\prime
}\right) ,\phi \left( {\bf r}^{\prime },\tau ^{\prime }\right) \right] &=&%
\frac{\hbar Sr_0^3}\epsilon \int d\tau ^{\prime }d^3{\bf r}^{\prime }\left\{
-i\epsilon ^{1/4}\sin \left( \theta _0+\sqrt{\epsilon }\overline{\delta }%
\right) \phi \left( \frac{\partial \overline{\delta }}{\partial \tau
^{\prime }}\right) \right.  \nonumber \\
&&+2\overline{K}_2^{\prime }\sin ^2\left( 2\phi \right) \sin ^4\left( \theta
_0+\sqrt{\epsilon }\overline{\delta }\right) +2\overline{H}_x\sin ^2\left( 
\frac \phi 2\right) \sin \left( \theta _0+\sqrt{\epsilon }\overline{\delta }%
\right)  \nonumber \\
&&+\frac 12\epsilon ^{3/2}\left( \nabla ^{\prime }\overline{\delta }\right)
^2+\frac 12\epsilon ^{1/2}\sin ^2\left( \theta _0+\sqrt{\epsilon }\overline{%
\delta }\right) \left( \nabla ^{\prime }\phi \right) ^2  \nonumber \\
&&\left. +\frac A4\epsilon ^{3/2}\left( \sqrt{6}\overline{\delta }^2-%
\overline{\delta }^3\right) \right\} ,  \eqnum{19}
\end{eqnarray}
where 
\begin{equation}
A=2\frac{\left| \cot \theta _H\right| ^{1/3}}{1+\left| \cot \theta _H\right|
^{2/3}}\left[ 1+\frac 43\left( \overline{K}_2-\overline{K}_2^{\prime
}\right) \frac{7-4\left| \cot \theta _H\right| ^{2/3}}{1+\left| \cot \theta
_H\right| ^{2/3}}\right] .  \eqnum{20}
\end{equation}
In Eq. (19) we have performed the integration by part for the first term and
have neglected the total imaginary-time derivative. In can be showed that
for $\pi /2<\theta _H<\pi $, only small values of $\phi $ contribute to the
path integral, so that one can replace $\sin ^2\phi $ in Eq. (19) by $\phi
^2 $ and neglect the term including $\left( \nabla ^{\prime }\phi \right) ^2$
which is of the order $\epsilon ^2$ while the other terms are of the order $%
\epsilon ^{3/2}$. Then the Gaussian integration over $\phi $ leads to 
\begin{equation}
\int {\cal D}\left\{ \delta \left( {\bf r}^{\prime },\tau ^{\prime }\right)
\right\} \exp \left( -\frac 1\hbar {\cal S}_E^{eff}\right) ,  \eqnum{21}
\end{equation}
where the effective action is 
\begin{equation}
{\cal S}_E^{eff}\left[ \overline{\delta }\left( {\bf r}^{\prime },\tau
^{\prime }\right) \right] =\hbar S\epsilon ^{1/2}r_0^3\int d\tau ^{\prime
}d^3{\bf r}^{\prime }\left[ \frac 12M\left( \frac{\partial \overline{\delta }%
}{\partial \tau ^{\prime }}\right) ^2+\frac 12\left( \nabla ^{\prime }%
\overline{\delta }\right) ^2+\frac A4\left( \sqrt{6}\overline{\delta }^2-%
\overline{\delta }^3\right) \right] .  \eqnum{22}
\end{equation}
The effect mass in Eq. (22) is found to be 
\begin{equation}
M=\frac{\left( 1+\left| \cot \theta _H\right| ^{2/3}\right) \left[ 1+\frac 83%
\left( \overline{K}_2-\overline{K}_2^{\prime }\right) \frac{\left| \cot
\theta _H\right| ^{2/3}}{1+\left| \cot \theta _H\right| ^{2/3}}\right] }{%
1-\epsilon +16\overline{K}_2^{\prime }+4\left( \overline{K}_2-\overline{K}%
_2^{\prime }\right) \frac 1{1+\left| \cot \theta _H\right| ^{2/3}}+128%
\overline{K}_2^{\prime }\left( \overline{K}_2-\overline{K}_2^{\prime
}\right) \frac{\left| \cot \theta _H\right| ^{2/3}}{1+\left| \cot \theta
_H\right| ^{2/3}}}.  \eqnum{23}
\end{equation}
Introducing the variables $\overline{\tau }=\tau ^{\prime }\sqrt{A/M}$ and $%
\overline{{\bf r}}={\bf r}^{\prime }\sqrt{A}$, the effective action Eq. (22)
is simplified as 
\begin{equation}
{\cal S}_E^{eff}\left[ \overline{\delta }\left( \overline{{\bf r}},\overline{%
\tau }\right) \right] =\hbar S\epsilon ^{1/2}r_0^3\frac{\sqrt{M}}A\int d%
\overline{\tau }d^3\overline{{\bf r}}\left[ \frac 12\left( \frac{\partial 
\overline{\delta }}{\partial \overline{\tau }}\right) ^2+\frac 12\left( 
\overline{\nabla }\overline{\delta }\right) ^2+\frac 14\left( \sqrt{6}%
\overline{\delta }^2-\overline{\delta }^3\right) \right] .  \eqnum{24}
\end{equation}

For the quantum reversal of magnetization ${\bf M}$ in a small particle of
volume $V\ll r_0^3$, ${\bf M}$ is uniform within the particle and $\overline{%
\delta }$ does not depend on the space $\overline{{\bf r}}$, Eq. (24)
reduces to 
\begin{equation}
{\cal S}_E^{eff}\left[ \overline{\delta }\left( \overline{{\bf r}},\overline{%
\tau }\right) \right] =\hbar S\epsilon ^{5/4}\sqrt{MA}V\int d\overline{\tau }%
\left[ \frac 12\left( \frac{d\overline{\delta }}{d\overline{\tau }}\right)
^2+\frac 14\left( \sqrt{6}\overline{\delta }^2-\overline{\delta }^3\right)
\right] .  \eqnum{25}
\end{equation}
The corresponding classical trajectory satisfies the equation of motion 
\begin{equation}
\frac{d^2\overline{\delta }}{d\overline{\tau }^2}=\frac 12\sqrt{6}\overline{%
\delta }-\frac 34\overline{\delta }^2.  \eqnum{26}
\end{equation}
Eq. (26) has the instanton solution 
\begin{equation}
\overline{\delta }\left( \overline{\tau }\right) =\frac{\sqrt{6}}{\cosh
^2\left( 3^{1/4}\times 2^{-5/4}\overline{\tau }\right) },  \eqnum{27}
\end{equation}
corresponding to the variation of $\overline{\delta }$ from $\overline{%
\delta }=0$ at $\overline{\tau }=-\infty $, to $\overline{\delta }=\sqrt{6}$
at $\overline{\tau }=0$, and then back to $\overline{\delta }=0$ at $%
\overline{\tau }=\infty $. Eq. (27) agrees well with the result in Refs. 13
and 15. The associated classical action is found to be 
\begin{eqnarray}
{\cal S}_{cl} &=&\frac{2^{17/4}\times 3^{1/4}}5\hbar S\epsilon ^{5/4} 
\nonumber \\
&&\times \frac{\left| \cot \theta _H\right| ^{1/6}\left[ 1+\frac 23\left( 
\overline{K}_2-\overline{K}_2^{\prime }\right) \frac{7-2\left| \cot \theta
_H\right| ^{2/3}}{1+\left| \cot \theta _H\right| ^{2/3}}\right] }{\sqrt{%
1-\epsilon +16\overline{K}_2^{\prime }+4\left( \overline{K}_2-\overline{K}%
_2^{\prime }\right) \frac 1{1+\left| \cot \theta _H\right| ^{2/3}}+128%
\overline{K}_2^{\prime }\left( \overline{K}_2-\overline{K}_2^{\prime
}\right) \frac{\left| \cot \theta _H\right| ^{2/3}}{1+\left| \cot \theta
_H\right| ^{2/3}}}}.  \eqnum{28}
\end{eqnarray}

Now we turn to the nonuniform problem. In case of a thin film of thickness $%
h $ less than the size $r_0/\epsilon ^{1/4}$ of the critical nucleus and its
plane is perpendicular to the initial easy axis, we obtain the action Eq.
(24) after performing the integration over the $\overline{z}$ variable, 
\begin{equation}
{\cal S}_E^{eff}\left[ \overline{\delta }\left( \overline{{\bf r}},\overline{%
\tau }\right) \right] =\hbar S\epsilon ^{3/4}r_0^2h\sqrt{\frac MA}\int d%
\overline{\tau }d^2\overline{{\bf r}}\left[ \frac 12\left( \frac{\partial 
\overline{\delta }}{\partial \overline{\tau }}\right) ^2+\frac 12\left( 
\overline{\nabla }\overline{\delta }\right) ^2+\frac 14\left( \sqrt{6}%
\overline{\delta }^2-\overline{\delta }^3\right) \right] .  \eqnum{29}
\end{equation}
At zero temperature the classical solution of the effective action Eq. (29)
has $O\left( 3\right) $ symmetry in two spatial plus one imaginary time
dimensions. Therefore, the solution $\overline{\delta }$ is a function of $u$%
, where $u=\left( \overline{\rho }^2+\overline{\tau }^2\right) ^{1/2}$, and $%
\overline{\rho }=\left( \overline{x}^2+\overline{y}^2\right) ^{1/2}$ is the
normalized distance from the ${\bf z}$ axis. Now the effective action Eq.
(29) becomes 
\begin{equation}
{\cal S}_E^{eff}\left[ \overline{\delta }\left( \overline{{\bf r}},\overline{%
\tau }\right) \right] =4\pi \hbar S\epsilon ^{3/4}r_0^2h\sqrt{\frac MA}\int
duu^2\left[ \frac 12\left( \frac{d\overline{\delta }}{du}\right) ^2+\frac 14%
\left( \sqrt{6}\overline{\delta }^2-\overline{\delta }^3\right) \right] . 
\eqnum{30}
\end{equation}
The corresponding classical trajectory satisfies the following equation of
motion 
\begin{equation}
\frac{d^2\overline{\delta }}{du^2}+\frac 2u\frac{d\overline{\delta }}{du}=%
\frac{\sqrt{6}}2\overline{\delta }-\frac 34\overline{\delta }^2.  \eqnum{31}
\end{equation}
By applying the similar method,\cite{4,6} the instanton solution of Eq. (31)
can be found numerically and is illustrated in Fig. 1. The maximal rotation
of ${\bf M}$ is $\overline{\delta }_{\max }\approx 6.8499$ at $\overline{%
\tau }=0$ and $\overline{\rho }=0$. Numerical integration in Eq. (30), using
this solution, gives the rate of quantum nucleation for a thin ferromagnetic
film as 
\begin{eqnarray}
\Gamma _Q &\varpropto &\exp \left( -{\cal S}_E/\hbar \right)  \nonumber \\
&=&\exp \left\{ -74.39S\epsilon ^{3/4}r_0^2h\frac{1+\left| \cot \theta
_H\right| ^{2/3}}{\left| \cot \theta _H\right| ^{1/6}}\left[ 1-\frac 23%
\left( \overline{K}_2-\overline{K}_2^{\prime }\right) \frac{7-6\left| \cot
\theta _H\right| ^{2/3}}{1+\left| \cot \theta _H\right| ^{2/3}}\right]
\right.  \nonumber \\
&&\left. \times \frac 1{\sqrt{1-\epsilon +16\overline{K}_2^{\prime }+4\left( 
\overline{K}_2-\overline{K}_2^{\prime }\right) \frac 1{1+\left| \cot \theta
_H\right| ^{2/3}}+128\overline{K}_2^{\prime }\left( \overline{K}_2-\overline{%
K}_2^{\prime }\right) \frac{\left| \cot \theta _H\right| ^{2/3}}{1+\left|
\cot \theta _H\right| ^{2/3}}}}\right\} .  \eqnum{32}
\end{eqnarray}

At high temperature, the nucleation of ${\bf M}$ is due to thermal
activation, and the rate of nucleation follows $\Gamma _T\varpropto \exp
\left( -W_{\min }/k_BT\right) $, where $W_{\min }$ is the minimal work
necessary to produce a nucleus capable of growing. In this case the
instanton solution becomes independent of the imaginary-time variable $%
\overline{\tau }$. In order to obtain $W_{\min }$, we consider the effective
potential of the system 
\begin{equation}
U_{eff}=\int d^3{\bf r}E=\int d^3{\bf r}\left[ \frac \alpha 2M_0^2\left(
\left( \nabla \theta \right) ^2+\sin ^2\theta \left( \nabla \phi \right)
^2\right) +E_a\left( \theta ,\phi \right) \right] .  \eqnum{33}
\end{equation}
For a cylindrical bubble Eq. (33) reduces to 
\begin{equation}
U_{eff}=4\pi K_1h\epsilon r_0^2\int_0^\infty d\overline{\rho }\overline{\rho 
}\left[ \frac 12\left( \frac{d\overline{\delta }}{d\overline{\rho }}\right)
^2+\frac 14\left( \sqrt{6}\overline{\delta }^2-\overline{\delta }^3\right)
\right] .  \eqnum{34}
\end{equation}
From the saddle point of the functional the shape of the critical nucleus
satisfies 
\begin{equation}
\frac{d^2\overline{\delta }}{d\overline{\rho }^2}+\frac 1{\overline{\rho }}%
\frac{d\overline{\delta }}{d\overline{\rho }}=\frac{\sqrt{6}}2\overline{%
\delta }-\frac 34\overline{\delta }^2.  \eqnum{35}
\end{equation}
The solution can be found by numerical method similar to the one in Refs. 4
and 6. Fig. 2 shows the shape of the critical bubble in thermal nucleation,
and the maximal size is $3.906$ at $\overline{\rho }=0$. Using this result,
the minimal work corresponding the thermal nucleation is 
\begin{equation}
W_{\min }=41.3376K_1h\epsilon r_0^2.  \eqnum{36}
\end{equation}
Comparing this with Eq. (32), we obtain the approximate formula for the
temperature characterizing the crossover from thermal to quantum nucleation
as 
\begin{eqnarray}
k_BT_c &\approx &0.55\frac{K_1\epsilon ^{1/4}}S\frac{\left| \cot \theta
_H\right| ^{1/6}}{1+\left| \cot \theta _H\right| ^{2/3}}\left[ 1+\frac 23%
\left( \overline{K}_2-\overline{K}_2^{\prime }\right) \frac{7-6\left| \cot
\theta _H\right| ^{2/3}}{1+\left| \cot \theta _H\right| ^{2/3}}\right] 
\nonumber \\
&&\times \left[ 1-\epsilon +16\overline{K}_2^{\prime }+4\left( \overline{K}%
_2-\overline{K}_2^{\prime }\right) \frac 1{1+\left| \cot \theta _H\right|
^{2/3}}\right.  \nonumber \\
&&\left. +128\overline{K}_2^{\prime }\left( \overline{K}_2-\overline{K}%
_2^{\prime }\right) \frac{\left| \cot \theta _H\right| ^{2/3}}{1+\left| \cot
\theta _H\right| ^{2/3}}\right] ^{1/2}.  \eqnum{37}
\end{eqnarray}
To observe the quantum nucleation one needs a large crossover temperature
and not too small a nucleation rate. Eq. (37) shows that ferromagnets with
large anisotropy, i.e., small ration of $K_2^{\prime }$ to $K_1$, and small
saturated magnetization are preferable for experimental study. In Fig. 3, we
plot the $\theta _H$ dependence of the crossover temperature $T_c$ for
typical values of parameters for nanometer-scale ferromagnets: $K_1=10^7$
erg/cm$^3$, $\overline{K}_2^{\prime }=0.1$, $\overline{K}_2-\overline{K}%
_2^{\prime }=0.01$, $M_0=500$ emu/cm$^3$, $\epsilon =0.01$ in a wide range
of angles $\pi /2<\theta _H<\pi $. Fig. 3 shows that the maximal value of $%
T_c$ is about 225 mK at $\theta _H=1.743$. The maximal value of $T_c$ as
well as $\Gamma _Q$ is expected to be observed in experiment. The similar $%
\theta _H$ dependence of the crossover temperature $T_c$ was first observed
in Ref. 15, while the problem considered in Ref. 15 was homogeneous spin
tunneling in single-domain particles with uniaxial symmetry.

\section*{IV. Ferromagnets with hexagonal symmetry}

In this section, we study the quantum nucleation of magnetization in
nanometer-scale ferromagnets with hexagonal symmetry in an external magnetic
field at an arbitrary angle in the $ZX$ plane. Now the magnetocrystalline
anisotropy energy $E_a\left( \theta ,\phi \right) $ can be written as 
\begin{eqnarray}
E_a\left( \theta ,\phi \right) &=&K_1\sin ^2\theta +K_2\sin ^4\theta
+K_3\sin ^6\theta -K_3^{\prime }\sin ^6\theta \cos \left( 6\phi \right) 
\nonumber \\
&&-M_0H_x\sin \theta \cos \phi -M_0H_z\cos \theta ,  \eqnum{38}
\end{eqnarray}
where $K_1$, $K_2$, $K_3$, and $K_3^{\prime }$ are the magnetic anisotropic
coefficients. The easy axes are $\pm \widehat{z}$ for $K_1>0$. By choosing $%
K_3^{\prime }>0$, we take $\phi =0$ to be the easy plane, at which the
anisotropy energy can be written in terms of the dimensionless parameters as 
\begin{equation}
\overline{E}_a\left( \theta ,\phi =0\right) =\frac 12\sin ^2\theta +%
\overline{K}_2\sin ^4\theta +\left( \overline{K}_3-\overline{K}_3^{\prime
}\right) \sin ^6\theta -\overline{H}\cos \left( \theta -\theta _H\right) +%
\overline{E}_0,  \eqnum{39}
\end{equation}
where $\overline{K}_3=K_3/2K_1$ and $\overline{K}_3^{\prime }=K_3^{\prime
}/2K_1$.

Then the initial angle $\theta _0$ is determined by $\left[ d\overline{E}%
_a\left( \theta ,0\right) /d\theta \right] _{\theta =\theta _0}=0$, and the
critical angle $\theta _c$ and the dimensionless critical field $\overline{H}%
_c$ by both $\left[ d\overline{E}_a\left( \theta ,0\right) /d\theta \right]
_{\theta =\theta _c,\overline{H}=\overline{H}_c}=0$ and $\left[ d^2\overline{%
E}_a\left( \theta ,0\right) /d\theta ^2\right] _{\theta =\theta _c,\overline{%
H}=\overline{H}_c}=0$, which leads to 
\begin{eqnarray}
&&\left. \frac 12\sin \left( 2\theta _0\right) +\overline{H}\sin \left(
\theta _0-\theta _H\right) +4\overline{K}_2\sin ^3\theta _0\cos \theta
_0+6\left( \overline{K}_3-\overline{K}_3^{\prime }\right) \sin ^5\theta
_0\cos \theta _0=0,\right.  \eqnum{40a} \\
&&\left. \frac 12\sin \left( 2\theta _c\right) +\overline{H}_c\sin \left(
\theta _c-\theta _H\right) +4\overline{K}_2\sin ^3\theta _c\cos \theta
_c+6\left( \overline{K}_3-\overline{K}_3^{\prime }\right) \sin ^5\theta
_c\cos \theta _c=0,\right.  \eqnum{40b} \\
&&\left. \cos \left( 2\theta _c\right) +\overline{H}_c\cos \left( \theta
_c-\theta _H\right) +4\overline{K}_2\left( 3\sin ^2\theta _c\cos ^2\theta
_c-\sin ^4\theta _c\right) \right.  \nonumber \\
&&\left. +6\left( \overline{K}_3-\overline{K}_3^{\prime }\right) \left(
5\sin ^4\theta _c\cos ^2\theta _c-\sin ^6\theta _c\right) =0,\right. 
\eqnum{40c}
\end{eqnarray}
Under the assumption that $\left| \overline{K}_2\right| $, $\left| \overline{%
K}_3-\overline{K}_3^{\prime }\right| \ll 1$, we obtain the dimensionless
critical field $\overline{H}_c$ and the critical angle as 
\begin{eqnarray}
\overline{H}_c &=&\frac 1{\left[ \left( \sin \theta _H\right) ^{2/3}+\left|
\cos \theta _H\right| ^{2/3}\right] ^{3/2}}\left[ 1+\frac{4\overline{K}_2}{%
1+\left| \cot \theta _H\right| ^{2/3}}+\frac{6\left( \overline{K}_3-%
\overline{K}_3^{\prime }\right) }{\left( 1+\left| \cot \theta _H\right|
^{2/3}\right) ^2}\right] ,  \eqnum{41a} \\
\sin \theta _c &=&\frac 1{\left( 1+\left| \cot \theta _H\right|
^{2/3}\right) ^{1/2}}\left[ 1+\frac 83\overline{K}_2\frac{\left| \cot \theta
_H\right| ^{2/3}}{1+\left| \cot \theta _H\right| ^{2/3}}\right.  \nonumber \\
&&\left. +8\left( \overline{K}_3-\overline{K}_3^{\prime }\right) \frac{%
\left| \cot \theta _H\right| ^{2/3}}{\left( 1+\left| \cot \theta _H\right|
^{2/3}\right) ^2}\right] .  \eqnum{41b}
\end{eqnarray}
In the limit of small $\epsilon =1-\overline{H}/\overline{H}_c$, Eq. (40a)
becomes 
\begin{eqnarray}
&&-\epsilon \overline{H}_c\sin \left( \theta _c-\theta _H\right) +\eta
^2\left[ \left( 3/2\right) \overline{H}_c\sin \left( \theta _c-\theta
_H\right) +3\overline{K}_2\sin \left( 4\theta _c\right) \right.  \nonumber \\
&&\left. +12\left( \overline{K}_3-\overline{K}_3^{\prime }\right) \sin
^3\theta _c\cos \theta _c\left( 5-8\sin ^2\theta _c\right) \right] +\eta
\left\{ \epsilon \overline{H}_c\cos \left( \theta _c-\theta _H\right) \right.
\nonumber \\
&&-\eta ^2\left. \left[ \left( 1/2\right) \overline{H}_c\cos \left( \theta
_c-\theta _H\right) +4\overline{K}_2\cos \left( 4\theta _c\right) \right.
\right.  \nonumber \\
&&\left. \left. \left. +12\left( \overline{K}_3-\overline{K}_3^{\prime
}\right) \sin ^2\theta _c\left( 5-20\sin ^2\theta _c+16\sin ^4\theta
_c\right) \right] \right\} =0,\right.  \eqnum{42}
\end{eqnarray}
where $\eta \equiv \theta _c-\theta _0$ which is small for $\epsilon \ll 1$.
By introducing a small variable $\delta \equiv \theta -\theta _0$ $\left(
\left| \delta \right| \ll 1\text{ in the limit of }\epsilon \ll 1\right) $,
the anisotropy energy becomes 
\begin{equation}
\overline{E}_a\left( \delta ,\phi \right) =\overline{K}_3^{\prime }\left[
1-\cos \left( 6\phi \right) \right] \sin ^6\left( \theta _0+\delta \right) +%
\overline{H}_x\left( 1-\cos \phi \right) \sin \left( \theta _0+\delta
\right) +\overline{E}_1\left( \delta \right) ,  \eqnum{43}
\end{equation}
where $\overline{E}_1\left( \delta \right) $ is a function of only $\delta $
given by 
\begin{eqnarray}
\overline{E}_1\left( \delta \right) &=&\left[ \frac 12\overline{H}_c\sin
\left( \theta _c-\theta _H\right) +\overline{K}_2\sin \left( 4\theta
_c\right) +4\left( \overline{K}_3-\overline{K}_3^{\prime }\right) \left(
5\sin ^3\theta _c\cos ^3\theta _c-3\sin ^5\theta _c\cos \theta _c\right)
\right]  \nonumber \\
&&\times \left( \delta ^3-3\delta ^2\eta \right) +\left[ \frac 18\overline{H}%
_c\cos \left( \theta _c-\theta _H\right) +\overline{K}_2\cos \left( 4\theta
_c\right) +3\left( \overline{K}_3-\overline{K}_3^{\prime }\right) \sin
^2\theta _c\left( \sin ^4\theta _c\right. \right.  \nonumber \\
&&\left. \left. -10\sin ^2\theta _c\cos ^2\theta _c+5\cos ^4\theta _c\right)
\right] \left( \delta ^4-4\delta ^3\eta +6\delta ^2\eta ^2-4\delta
^2\epsilon \right) +\epsilon \delta ^2\left[ 4\overline{K}_2\cos \left(
4\theta _c\right) \right.  \nonumber \\
&&\left. +12\left( \overline{K}_3-\overline{K}_3^{\prime }\right) \sin
^2\theta _c\left( \sin ^4\theta _c-10\sin ^2\theta _c\cos ^2\theta _c+5\cos
^4\theta _c\right) \right] .  \eqnum{44}
\end{eqnarray}

By applying the similar procedure in Sec. III, we obtain the transition
amplitude Eqs. (21) and (22) by integrating out $\phi $. For this case the
effective mass is 
\begin{eqnarray}
M &=&\left( 1+\left| \cot \theta _H\right| ^{2/3}\right) \left[ 1+\frac 83%
\overline{K}_2\frac{\left| \cot \theta _H\right| ^{2/3}}{1+\left| \cot
\theta _H\right| ^{2/3}}+8\left( \overline{K}_3-\overline{K}_3^{\prime
}\right) \frac{\left| \cot \theta _H\right| ^{2/3}}{\left( 1+\left| \cot
\theta _H\right| ^{2/3}\right) ^2}\right]  \nonumber \\
&&\times \left[ 1-\epsilon +36\overline{K}_3^{\prime }\frac 1{1+\left| \cot
\theta _H\right| ^{2/3}}+4\overline{K}_2\left( 1+120\overline{K}_3^{\prime }%
\frac{\left| \cot \theta _H\right| ^{2/3}}{1+\left| \cot \theta _H\right|
^{2/3}}\right) \frac 1{1+\left| \cot \theta _H\right| ^{2/3}}\right. 
\nonumber \\
&&\left. +6\left( \overline{K}_3-\overline{K}_3^{\prime }\right) \left( 1+240%
\overline{K}_3^{\prime }\frac{\left| \cot \theta _H\right| ^{2/3}}{1+\left|
\cot \theta _H\right| ^{2/3}}\right) \frac 1{\left( 1+\left| \cot \theta
_H\right| ^{2/3}\right) ^2}\right] ^{-1},  \eqnum{45}
\end{eqnarray}
and the prefactor $A$ is 
\begin{equation}
A=2\frac{\left| \cot \theta _H\right| ^{1/3}}{1+\left| \cot \theta _H\right|
^{2/3}}\left[ 1+\frac 43\overline{K}_2\frac{7-4\left| \cot \theta _H\right|
^{2/3}}{1+\left| \cot \theta _H\right| ^{2/3}}+2\left( \overline{K}_3-%
\overline{K}_3^{\prime }\right) \frac{11-16\left| \cot \theta _H\right|
^{2/3}}{\left( 1+\left| \cot \theta _H\right| ^{2/3}\right) ^2}\right] . 
\eqnum{46}
\end{equation}
In case of a small ferromagnet of volume $V\ll r_0^3$, the result of quantum
nucleation is $\Gamma _Q\varpropto \exp \left( -{\cal S}_{cl}/\hbar \right) $%
, where the classical action for hexagonal symmetry is found to be 
\begin{eqnarray}
{\cal S}_{cl} &=&\frac{2^{17/4}\times 3^{1/4}}5\hbar S\epsilon ^{5/4}\left|
\cot \theta _H\right| ^{1/6}  \nonumber \\
&&\times \left[ 1+\frac 23\overline{K}_2\frac{7-2\left| \cot \theta
_H\right| ^{2/3}}{1+\left| \cot \theta _H\right| ^{2/3}}+\left( \overline{K}%
_3-\overline{K}_3^{\prime }\right) \frac{11-12\left| \cot \theta _H\right|
^{2/3}}{\left( 1+\left| \cot \theta _H\right| ^{2/3}\right) ^2}\right] 
\nonumber \\
&&\times \left[ 1-\epsilon +36\overline{K}_3^{\prime }\frac 1{1+\left| \cot
\theta _H\right| ^{2/3}}+4\overline{K}_2\left( 1+120\overline{K}_3^{\prime }%
\frac{\left| \cot \theta _H\right| ^{2/3}}{1+\left| \cot \theta _H\right|
^{2/3}}\right) \frac 1{1+\left| \cot \theta _H\right| ^{2/3}}\right. 
\nonumber \\
&&\left. +6\left( \overline{K}_3-\overline{K}_3^{\prime }\right) \left( 1+240%
\overline{K}_3^{\prime }\frac{\left| \cot \theta _H\right| ^{2/3}}{1+\left|
\cot \theta _H\right| ^{2/3}}\right) \frac 1{\left( 1+\left| \cot \theta
_H\right| ^{2/3}\right) ^2}\right] ^{-1/2},  \eqnum{47}
\end{eqnarray}
In case of a thin film of thickness $h$ less than the size $r_0/\epsilon
^{1/4}$ of the critical nucleus, we obtain the quantum nucleation as $\Gamma
_Q\varpropto \exp \left( -{\cal S}_E/\hbar \right) $, with the classical
action 
\begin{eqnarray}
{\cal S}_E &=&74.39S\epsilon ^{3/4}r_0^2h\frac{1+\left| \cot \theta
_H\right| ^{2/3}}{\left| \cot \theta _H\right| ^{1/6}}  \nonumber \\
&&\times \left[ 1-\frac 23\overline{K}_2\frac{7-2\left| \cot \theta
_H\right| ^{2/3}}{1+\left| \cot \theta _H\right| ^{2/3}}-\left( \overline{K}%
_3-\overline{K}_3^{\prime }\right) \frac{11-12\left| \cot \theta _H\right|
^{2/3}}{\left( 1+\left| \cot \theta _H\right| ^{2/3}\right) ^2}\right] 
\nonumber \\
&&\times \left[ 1-\epsilon +36\overline{K}_3^{\prime }\frac 1{1+\left| \cot
\theta _H\right| ^{2/3}}+4\overline{K}_2\left( 1+120\overline{K}_3^{\prime }%
\frac{\left| \cot \theta _H\right| ^{2/3}}{1+\left| \cot \theta _H\right|
^{2/3}}\right) \frac 1{1+\left| \cot \theta _H\right| ^{2/3}}\right. 
\nonumber \\
&&\left. +6\left( \overline{K}_3-\overline{K}_3^{\prime }\right) \left( 1+240%
\overline{K}_3^{\prime }\frac{\left| \cot \theta _H\right| ^{2/3}}{1+\left|
\cot \theta _H\right| ^{2/3}}\right) \frac 1{\left( 1+\left| \cot \theta
_H\right| ^{2/3}\right) ^2}\right] ^{-1/2}.  \eqnum{48}
\end{eqnarray}
And the crossover temperature for hexagonal symmetry is found to be 
\begin{eqnarray}
k_BT_c &\approx &0.55\frac{K_1\epsilon ^{1/4}}S\frac{\left| \cot \theta
_H\right| ^{1/6}}{1+\left| \cot \theta _H\right| ^{2/3}}  \nonumber \\
&&\times \left[ 1+\frac 23\overline{K}_2\frac{7-2\left| \cot \theta
_H\right| ^{2/3}}{1+\left| \cot \theta _H\right| ^{2/3}}+\left( \overline{K}%
_3-\overline{K}_3^{\prime }\right) \frac{11-12\left| \cot \theta _H\right|
^{2/3}}{\left( 1+\left| \cot \theta _H\right| ^{2/3}\right) ^2}\right] 
\nonumber \\
&&\times \left[ 1-\epsilon +36\overline{K}_3^{\prime }\frac 1{1+\left| \cot
\theta _H\right| ^{2/3}}+4\overline{K}_2\left( 1+120\overline{K}_3^{\prime }%
\frac{\left| \cot \theta _H\right| ^{2/3}}{1+\left| \cot \theta _H\right|
^{2/3}}\right) \frac 1{1+\left| \cot \theta _H\right| ^{2/3}}\right. 
\nonumber \\
&&\left. +6\left( \overline{K}_3-\overline{K}_3^{\prime }\right) \left( 1+240%
\overline{K}_3^{\prime }\frac{\left| \cot \theta _H\right| ^{2/3}}{1+\left|
\cot \theta _H\right| ^{2/3}}\right) \frac 1{\left( 1+\left| \cot \theta
_H\right| ^{2/3}\right) ^2}\right] ^{1/2}.  \eqnum{49}
\end{eqnarray}

\section*{V. Conclusions and discussions}

In summary we have investigated the quantum nucleation of magnetization in
nanometer-scale ferromagnets in the presence of an external magnetic field
at arbitrary angle. We consider the magnetocrystalline anisotropy with
tetragonal symmetry and that with hexagonal symmetry, respectively. By
applying the instanton method in the spin-coherent-state path-integral
representation, we obtain the analytical formulas for quantum reversal of
magnetization in small magnets and the numerical formulas for quantum
nucleation in thin ferromagnetic film in a wide range of angles $\pi
/2<\theta _H<\pi $. The temperature characterizing the crossover from the
quantum to thermal nucleation is clearly shown for each case. Our results
show that the rate of quantum nucleation and the crossover temperature
depend on the orientation of the external magnetic field distinctly.
Therefore, both the orientation and the strength of the external magnetic
field are the controllable parameters for the experimental test of quantum
nucleation of magnetization in nanometer-scale ferromagnets. If the
experiment is to be performed, there are three control parameters for
comparison with theory: the angle of the external magnetic field $\theta _H$%
, the strength of the field in terms of $\epsilon $, and the temperature $T$%
. Our results show that ferromagnetic samples with large anisotropy and
small saturated magnetization are suitable for experimental study the
phenomenon of quantum nucleation.

Recently, Wernsdorfer and co-workers have performed the switching field
measurements on individual ferrimagnetic and insulating BaFeCoTiO
nanoparticles containing about $10^5$-$10^6$ spins at very low temperatures
(0.1-6K).\cite{14} They found that above 0.4K, the magnetization reversal of
these particles is unambiguously described by the N\'{e}el-Brown theory of
thermal activated rotation of the particle's moment over a well defined
anisotropy energy barrier. Below 0.4K, strong deviations from this model are
evidenced which are quantitatively in agreement with the predictions of the
MQT theory without dissipation.\cite{13,15} It is noted that the observation
of quantum nucleation in ferromagnets would be extremely interesting as the
next example, after single-domain nanoparticles, of macroscopic quantum
tunneling. The theoretical results presented here may be useful for checking
the general theory in a wide range of systems, with more general symmetries.
The experimental procedures on single-domain ferromagnetic\ nanoparticles of
Barium ferrite with uniaxial symmetry\cite{14} may be applied to the systems
with more general symmetries. Note that the inverse of the WKB exponent $%
B^{-1}$ is the magnetic viscosity $S$ at the quantum-tunneling-dominated
regime $T\ll T_c$ studied by magnetic relaxation measurements.\cite{1}
Therefore, the quantum nucleation of magnetization should be checked at any $%
\theta _H$ by magnetic relaxation measurements. Over the past years a lot of
experimental and theoretical works were performed on the spin tunneling in
molecular Mn$_{12}$-Ac\cite{16} and Fe$_8$\cite{17} clusters having a
collective spin state $S=10$ (in this paper $S=10^6$). Further experiments
should focus on the level quantization of collective spin states of $S=10^2$-%
$10^4$.

The ferromagnet is typically an insulating particle with as many as $%
10^3\sim 10^6$ magnetic moments. For the dynamical process, it is important
to include the effect of the environment on quantum tunneling caused by
phonons, \cite{18,19} nucleation spins,\cite{20} and Stoner excitations and
eddy currents in metallic magnets.\cite{21} However, many studies showed
that even though these couplings might be crucial in macroscopic quantum
coherence, they are not strong enough to make the quantum tunneling
unobservable.\cite{1,18,19,20,21} The theoretical calculations performed in
this paper can be extended to the AFM\ bubbles, where the relevant quantity
is the excess spin due to the small noncompensation of two sublattices. Work
along this line is still in progress. We hope that the theoretical results
presented in this paper may stimulate more experiments whose aim is
observing quantum nucleation in nanometer-scale ferromagnets.

\section*{Acknowledgments}

Y. Zhou and R.L. would like to acknowledge Dr. Su-Peng Kou, Professor Zhan
Xu, Professor Mo-Lin Ge, Professor Jiu-Qing Liang and Professor Fu-Cho Pu
for stimulating discussions. R. L. would like to thank Professor W.
Wernsdorfer and Professor R. Sessoli for providing their paper (Ref. 11),
and Professor Kim for providing his paper (Ref. 13).

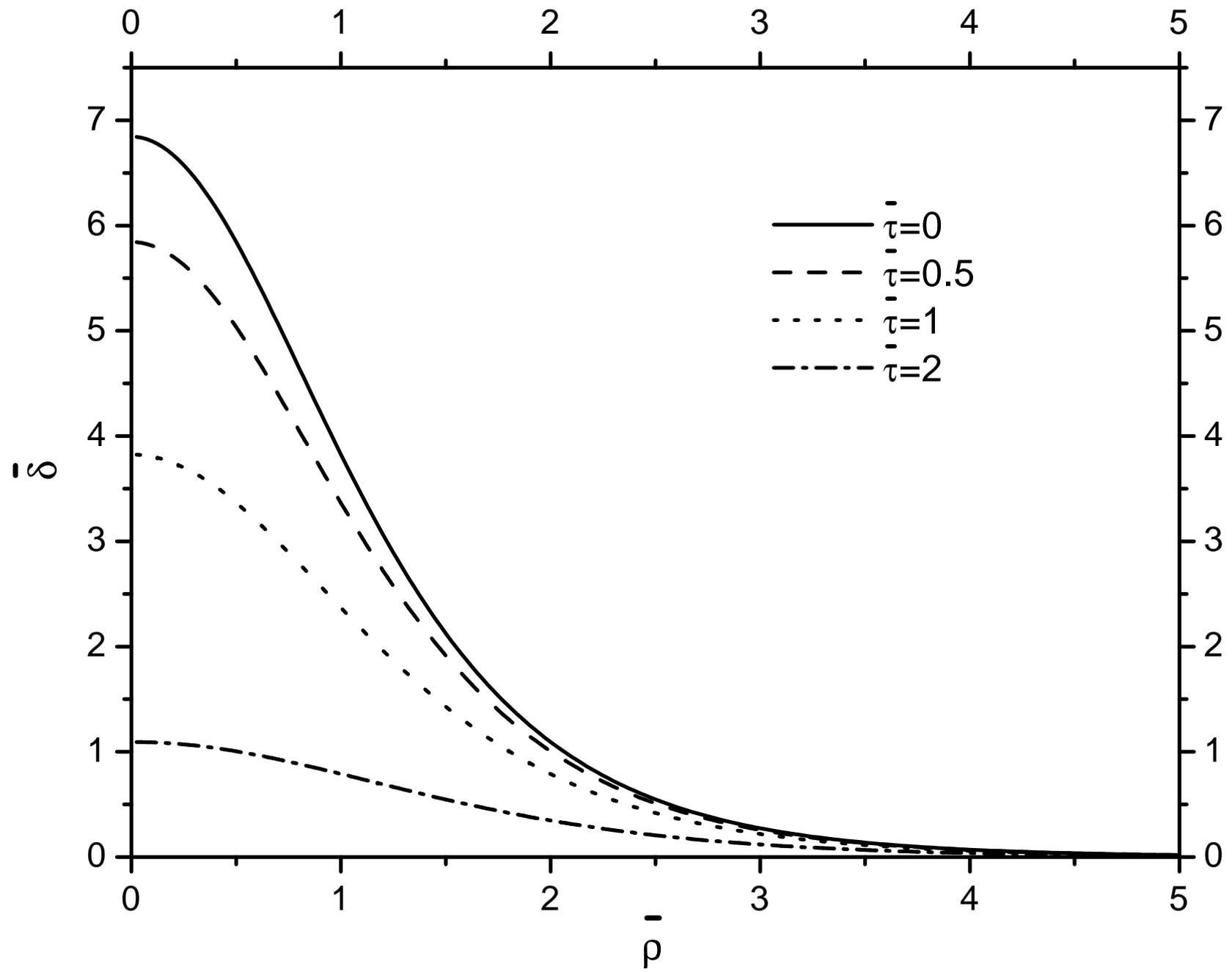
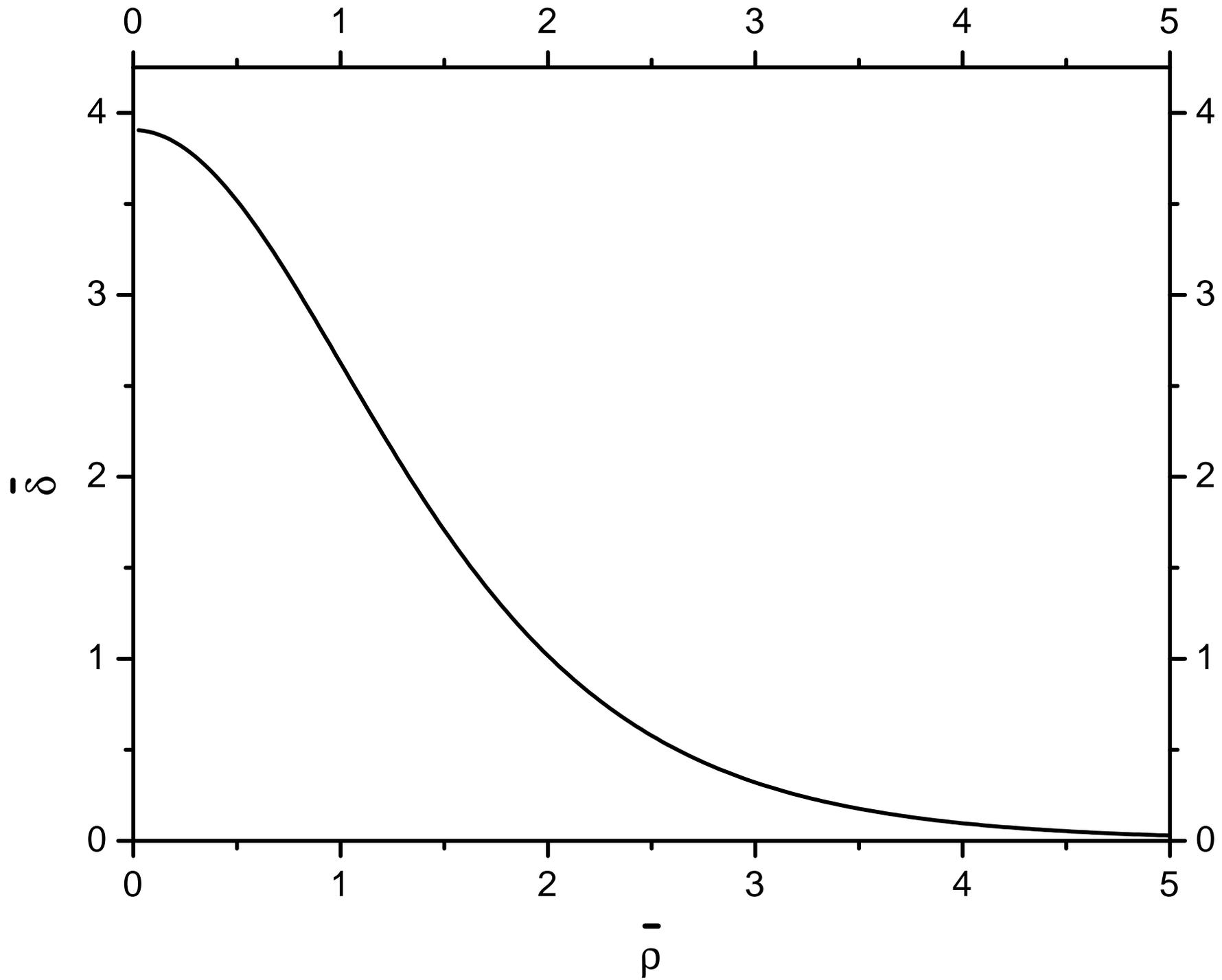
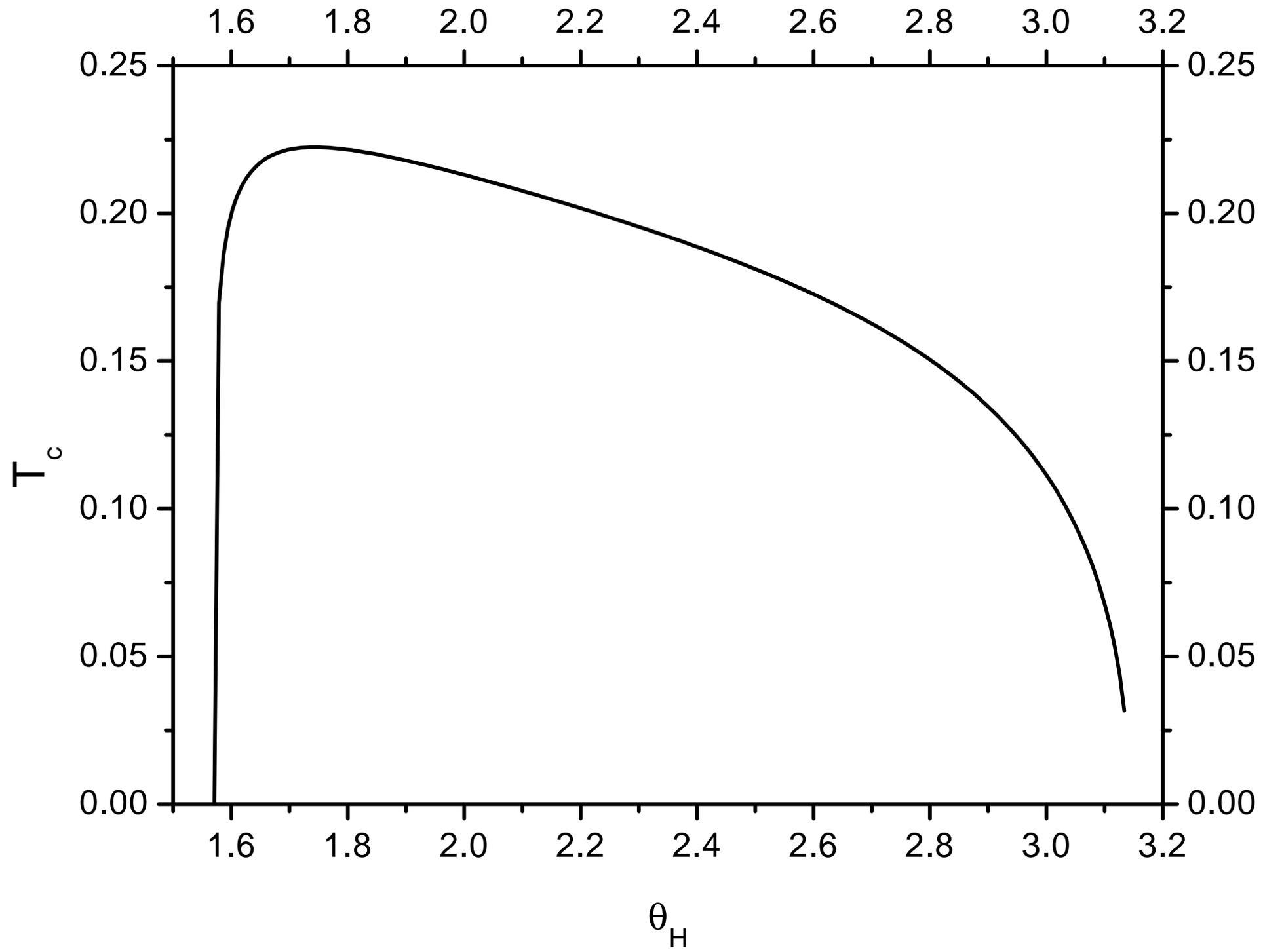
\end{document}